\documentclass[preprint, 5p]{elsarticle}

\usepackage{amsmath,amssymb,bm,color}
\usepackage{lineno,hyperref}
\usepackage{multirow}
\usepackage{float}
\usepackage{booktabs}
\usepackage{array}
\usepackage{epstopdf}

\begin{document}

\begin{frontmatter}

\title{Microscopic description of the proton halo in $^{12}$N}

\author[INPC]{K. Y. Zhang\corref{cor1}}
\ead{zhangky@caep.cn}

\author[INPC]{X. X. Lu\corref{cor1}}
\ead{lxx5187@qq.com}

\date{\today}

\cortext[cor1]{Corresponding Author}

\address[INPC]{Institute of Nuclear Physics and Chemistry, China Academy of Engineering Physics, Mianyang, Sichuan 621900, China}

\begin{abstract}

The year 2025 marks the 40th anniversary of the discovery of halo nuclei and the 15th anniversary of the development of the deformed relativistic Hartree-Bogoliubov theory in continuum (DRHBc). In this work, we present the first DRHBc description of the proton halo phenomenon. The available experimental proton separation energies and empirical matter root-mean-square (rms) radii are reasonably well reproduced for the $N=5$ isotones, ranging from the stable nucleus $^{9}$Be to the drip-line nucleus $^{12}$N. In particular, the DRHBc theory captures the abrupt increase in rms radii at $^{12}$N, unambiguously corroborating its proton halo structure. The formation of this halo is attributed to the occupation of a weakly bound orbital with dominant $1p$ components by the valence proton, which contributes approximately $90\%$ to the diffused proton density of $^{12}$N at large distances from the nuclear center. A shape decoupling between the prolate core and the oblate halo in $^{12}$N is predicted.

\end{abstract}

\end{frontmatter}


\section{Introduction}

In 1985, an anomalously large reaction cross section was observed in the $^{11}$Li nucleus \cite{Tanihata1985PRL}, later identified as a manifestation of the nuclear halo \cite{Hansen1987EPL,Kobayashi1988PRL}, a phenomenon that posed serious challenges to conventional nuclear theory. This discovery catalyzed the development of new-generation rare isotope beam facilities worldwide, significantly broadening the scope of nuclear physics research. Over the subsequent four decades, a range of exotic phenomena in nuclei near the dripline have been uncovered, such as the disappearance of traditional magic numbers and the emergence of new ones \cite{Ozawa2000PRL}. Experimentally, about 20 halo nuclei or candidates have been identified or proposed; see Fig. 1.4 of Ref. \cite{Tanihata2013PPNP} and Fig. 1 of Ref. \cite{Zhang2023PRC(L1)}.

On the theoretical side, considerable efforts have been devoted to describing known halo phenomena and predicting new halo candidates using approaches including the few-body model \cite{Zhukov1993PhysRep}, shell model \cite{Otsuka1993PRL}, antisymmetrized molecular dynamics \cite{Horiuchi1994ZPA}, halo effective field theory \cite{Ryberg2014PRC,Hammer2017JPG}, \textit{ab initio} calculations within the no-core shell model with continuum \cite{Calci2016PRL} and nuclear lattice effective field theory \cite{Shen2025PRL}, and nonrelativistic \cite{Schunck2008PRC} as well as relativistic \cite{Ring1996PPNP} density functional theories. Within the relativistic density functional framework, the relativistic continuum Hartree-Bogoliubov (RCHB) theory was established in 1996, achieving a microscopic, self-consistent description of the neutron halo in $^{11}$Li \cite{Meng1996PRL}. A novel phenomenon termed the \textit{giant halo}, comprising more than two neutrons, was predicted by the RCHB theory \cite{Meng1998PRL}. In 2010, the deformed relativistic Hartree-Bogoliubov theory in continuum (DRHBc), a deformed extension of the RCHB theory, was developed \cite{Zhou2010PRC(R)}. Over the past 15 years, the DRHBc theory has successfully described halo phenomena in $^{17,19}$B \cite{Yang2021PRL,Sun2021PRC(1)}, $^{15,19,22}$C \cite{Sun2018PLB,Sun2020NPA,Wang2024EPJA}, $^{31}$Ne \cite{Zhong2022SciChina,Pan2024PLB}, and the heaviest known halo nucleus, $^{37}$Mg \cite{Zhang2023PLB,An2024PLB}. It has also predicted deformed halo candidates such as $^{39}$Na \cite{Zhang2023PRC(L1)} and $^{42,44}$Mg \cite{Zhou2010PRC(R),Li2012PRC}. A triaxial extension, the triaxial relativistic Hartree-Bogoliubov theory in continuum (TRHBc), has recently been developed and applied to explore possible halos in triaxially deformed nuclei \cite{Zhang2023PRC(L2),Zhang2025arXiv} and to investigate other nuclear phenomena \cite{Lu2024PLB,Huang2025PRC}. Notably, an international collaboration has been constructing a global nuclear mass table based on the DRHBc theory~\cite{Zhang2020PRC,Pan2022PRC,Zhang2022ADNDT,Guo2024ADNDT}. For more relevant studies based on the DRHBc theory, see the website for the DRHBc mass table~\cite{DRHBc} and recent reviews~\cite{Sun2024NPR,Zhang2024NPR,Zhang2025AAPPS}.

The DRHBc descriptions and predictions of halo nuclei have so far been limited to neutron halos. Due to the additional confining effect from the Coulomb barrier, the formation of a proton halo is more difficult \cite{Riisager1992NPA}, and the number of known or proposed proton halo nuclei remains significantly smaller than that of neutron halo ones \cite{Tanihata2013PPNP,Zhang2023PRC(L1)}. A recent case of intense debate concerns the possibility of a proton halo in $^{22}$Al \cite{Lee2020PRL,Yu2024PRL}, which has been examined using the DRHBc and TRHBc theories \cite{Zhang2024PRC,Panagiota2025arXiv}. While it is now generally accepted that the ground state of $^{22}$Al does not exhibit a halo structure \cite{Sun2024CPC}, whether its proton-unbound excited state qualifies as a halo state remains a controversial question. Notably, excited states that have been proposed to exhibit halo characteristics--such as in $^6$Li \cite{Li2002PLB}, $^{10}$Be \cite{Al-Khalili2006PRC}, and $^{17}$F \cite{Morlock1997PRL}--are all particle-bound.

In this work, we present the first DRHBc description of the proton halo phenomenon, using the candidate nucleus $^{12}$N as an illustrative example. Early experimental evidence for a proton halo in $^{12}$N dates back to 1998, when Warner et al. reported an enhanced total reaction cross section on a Si target for $^{12}$N compared to its neighboring nuclei \cite{Warner1998NPA}. Further support came in 2006, when a subsequent study led by Warner et al. observed a large one-proton removal cross section for $^{12}$N, reinforcing the possibility of a proton halo \cite{Warner2006PRC}. In 2010, Li et al. remeasured the total reaction cross section of $^{12}$N in Si and interpreted the results as indicative of an extended proton density distribution \cite{Li2010CPL}. Consequently, $^{12}$N was classified as a proton halo nucleus in the 2013 review by Tanihata et al. \cite{Tanihata2013PPNP}.

While the halo structure may have significant implications for nuclear reaction observables, as demonstrated by a recent microscopic study of fusion reactions \cite{Sun2023PRC}, the $^{12}$N nucleus is also of particular astrophysical interest due to the critical reaction $^{11}$C($p,\gamma$)$^{12}$N that contributes to the production of CNO seed nuclei \cite{Tang2003PRC,Timofeyuk2003NPA}.
Another important process is $^{12}$C($\nu_e,e^-$)$^{12}$N, which plays a crucial role in probing the weak hadronic current and in neutrino detection using liquid scintillator detectors \cite{Maschuw1998PPNP}.

Although $^{12}$N has been involved in several earlier theoretical studies, its halo structure and density distribution were not examined within the \textit{ab initio} no-core shell model \cite{Navratil2003PRC}. Calculations based on the phenomenological cluster-core model \cite{Gupta2002JPG} and single-particle model assuming Woods-Saxon (WS) potentials \cite{Typel2005NPA} supported a $^{11}$C$+p$ configuration. A Skyrme Hartree-Fock study with the particle-number-conserving method for pairing assumed spherical nuclear shape and did not analyze proton densities \cite{Bai1997PRC}. In relativistic mean-field calculations, deformation was included but pairing correlations were neglected \cite{Zhou2000CPL,Chen2006IJMPE}. Therefore, a microscopic description of $^{12}$N based on the DRHBc theory--incorporating deformation, pairing, and continuum effects in a fully self-consistent manner--is both timely and essential.

\section{Theoretical framework}\label{theory}

In the DRHBc theory, the relativistic Hartree-Bogoliubov equations for quasiparticles \cite{Kucharek1991ZPA} read
\begin{equation}\label{RHB}
\left(\begin{matrix}
h_D-\lambda & \Delta \\
-\Delta^* &-h_D^*+\lambda
\end{matrix}\right)\left(\begin{matrix}
U_k\\
V_k
\end{matrix}\right)=E_k\left(\begin{matrix}
U_k\\
V_k
\end{matrix}\right),
\end{equation}
where $\lambda$ is the Fermi energy, $h_D$ denotes the Dirac Hamiltonian,
\begin{equation}
      h_D(\bm{r})=\bm{\alpha}\cdot\bm{p}+V(\bm{r})+\beta[M+S(\bm{r})],
\end{equation}
with $S(\bm r)$ and $V(\bm r)$ being the scalar and vector potentials, respectively, and $\Delta$ represents the pairing potential,
\begin{equation}\label{Delta}
\Delta(\bm r_1,\bm r_2) = V^{\mathrm{pp}}(\bm r_1,\bm r_2)\kappa(\bm r_1,\bm r_2),
\end{equation}
with the paring force $V^{\mathrm{pp}}$ and the paring tensor $\kappa$ \cite{Peter1980Book}.
$E_k$ and $(U_k, V_k)^{\rm T}$ are the quasiparticle energy and wave function, respectively.

To incorporate the large spatial extension of halo nuclei, the quasiparticle wave functions are expanded in a set of Dirac WS (DWS) basis \cite{Zhou2003PRC,Zhang2022PRC}.
This basis is obtained by solving the Dirac equation of spherical WS potentials and provides appropriate asymptotic behavior at large $r$.
It turns out that the solution in the DWS basis is essentially equivalent to the coordinate-space solution \cite{Zhou2003PRC} and more suitable for the description of weakly bound nuclei than the harmonic oscillator basis~\cite{Zhou2000CPL}.

For axially deformed nuclei, the densities and potentials can be expanded in terms of the Legendre polynomials,
\begin{equation}\label{legendre}
f(\bm r)=\sum_l f_l(r)P_l(\cos\theta),
\end{equation}
where the expansion order $l$ is limited to even numbers if spatial reflection symmetry is assumed \cite{Xiang2023Symmetry}.
This expansion can be generalized to spherical harmonics,
\begin{equation}
    f(\bm r)=\sum_{lm} f_{lm}(r)Y_{lm}(\Omega),
\end{equation}
enabling the inclusion of triaxial deformation within the TRHBc framework \cite{Zhang2023PRC(L2),Zhang2025arXiv}. We note that the nuclei studied in this work, $^{12}$N and its neighboring $N=5$ isotones, exhibit axial deformation in their ground states according to our TRHBc calculations.

\section{Results and discussion}\label{results}

The numerical details in our DRHBc calculations follow those in Ref. \cite{Li2012PRC}, which presents the first detailed formalism and numerical implementation of the DRHBc theory. Specifically, a density-dependent zero-range pairing force \cite{Meng1998PRC} is employed, with a pairing window of 60 MeV. The DWS basis is constructed with a box size of 20 fm, a mesh size of 0.1 fm, a cutoff energy of 100 MeV, and an angular momentum cutoff of $19/2$ $\hbar$. The number of negative-energy (Dirac sea) states is taken to be equal to that of the positive-energy (Fermi sea) states. The Legendre expansion \eqref{legendre} is truncated at $l_{\mathrm{max}} = 4$. All these numerical details have been tested to ensure convergence and physical reliability of the results \cite{Li2012PRC}. No parameter fitting is involved in this work.

\begin{figure}[htbp]
  \centering
  \includegraphics[width=0.4\textwidth]{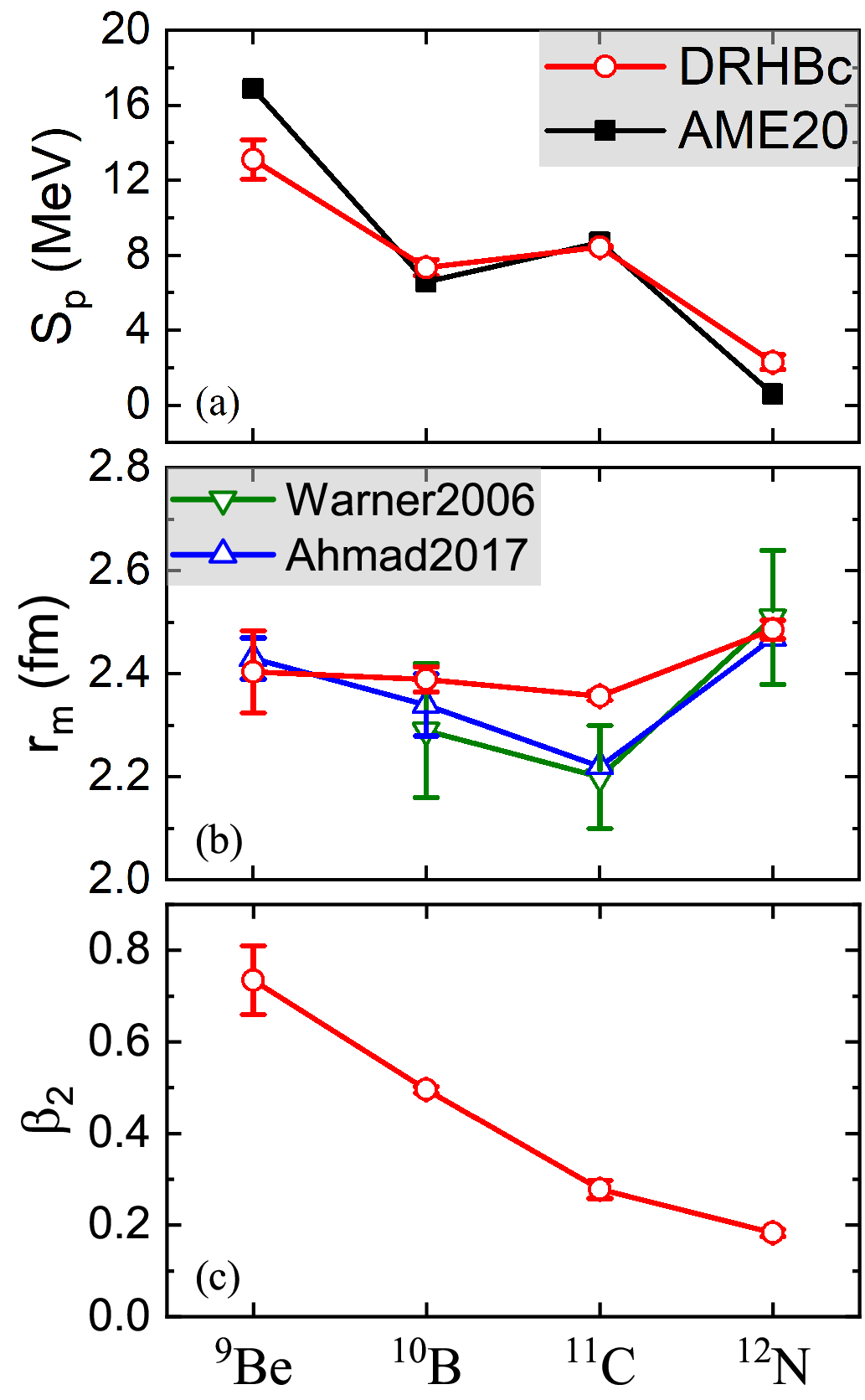}
  \caption{(a) Proton separation energies, (b) matter rms radii, and (c) quadrupole deformation parameters for the $N=5$ isotones from $Z=4$ ($^9$Be) to $Z=7$ ($^{12}$N). In order to quantify theoretical uncertainties, in the DRHBc calculations the density functionals NL3$^*$ \cite{Lalazissis2009PLB}, NL3 \cite{Lalazissis1997PRC}, NLSH \cite{Sharma1993PLB}, and PK1 \cite{Long2004PRC} are employed, with the pairing strength varied from $-340$ to $-380$~MeV~fm$^3$. For comparison, the experimental separation energies from AME20 \cite{AME2020(3)} are shown in (a), and the empirical matter radii estimated by Warner et al. \cite{Warner2006PRC} and Ahmad et al. \cite{Ahmad2017PRC} are shown in (b).}
\label{fig1}
\end{figure}

The DRHBc calculated proton separation energies ($S_p$), matter root-mean-square (rms) radii ($r_m$), and quadrupole deformation parameters ($\beta_2$) for the $N=5$ isotones, $^9$Be, $^{10}$B, $^{11}$C, and $^{12}$N, are shown in Fig. \ref{fig1}.
In modern nuclear structure studies, it is common practice to estimate theoretical uncertainties arising from the choice of effective interactions, e.g. by performing calculations with different parameter sets. To quantify the uncertainties in the DRHBc results, we have conducted a series of calculations using four widely employed density functionals: NL3$^*$ \cite{Lalazissis2009PLB}, NL3 \cite{Lalazissis1997PRC}, NLSH \cite{Sharma1993PLB}, and PK1 \cite{Long2004PRC}, with the pairing strength varied from $-340$ \cite{Xia2018ADNDT} to $-380$~MeV~fm$^3$ \cite{Zhou2010PRC(R)}.
Available experimental data for $S_p$ from AME20 \cite{AME2020(3)} and empirical $r_m$ values from Refs. \cite{Warner2006PRC,Ahmad2017PRC} are also included in Fig.~\ref{fig1} for comparison.
In Fig. \ref{fig1}(a), the DRHBc theory reproduces the proton separation energies reasonably well for $^{10}$B and $^{11}$C, while that of the proton drip-line nucleus $^{12}$N is slightly overestimated.
We note that the DRHBc calculations generally reproduce the experimental binding energy of $^{9}$Be with an accuracy of $\lesssim1$~MeV, and the underestimation of its proton separation energy originates from an overestimation of the experimental binding energy of $^{8}$Li.
In Fig. \ref{fig1}(b), good agreement within uncertainties is also observed, except for $^{11}$C.
It should be noted that the empirical $r_m$ values were extracted from reaction cross sections using the Glauber model \cite{Warner2006PRC,Ahmad2017PRC}, and are therefore not fully model independent.
The DRHBc results correctly reproduce the decreasing trend in $r_m$ from $^9$Be to $^{11}$C, which is consistent with the reduction in $\beta_2$ shown in Fig. \ref{fig1}(c).
However, the sudden increase in $r_m$ at $^{12}$N--captured by the DRHBc theory in both trend and magnitude--contrasts with the continued decrease in deformation, clearly indicating the emergence of a proton halo structure in $^{12}$N.

We note that the electric quadrupole and magnetic dipole moments of the nuclei studied in this work have been experimentally measured \cite{Stone2005ADNDT, Stone2016ADNDT}. It certainly makes sense to calculate these electromagnetic moments and compare them with experimental data. However, these observables are defined in the laboratory frame and differ from the expectation values obtained using intrinsic wave functions that break symmetries of the nuclear Hamiltonian \cite{Peter1980Book}. An adequate description requires going beyond the mean-field approximation and restoring the broken symmetries--for instance, through angular momentum projection (AMP) techniques \cite{Bender2003RMP, Niksic2011PPNP, Egido2016PS, Sheikh2021JPG}, which also enable the calculation of low-lying spectra. In this context, AMP has already been implemented within the DRHBc framework for even-even nuclei, successfully reproducing the ground-state rotational bands of $^{36,38,40}$Mg and providing insights into rotational modes of the deformed halo nuclei $^{42,44}$Mg \cite{Sun2021PRC(2), Sun2021SciBull}. It is therefore highly desirable to extend the DRHBc + AMP approach to odd-mass and odd-odd nuclei in the future and use this approach to compute electromagnetic moments and spectroscopic properties for the nuclei under investigation in this work.

\begin{figure*}[htbp]
  \centering
  \includegraphics[width=1.0\textwidth]{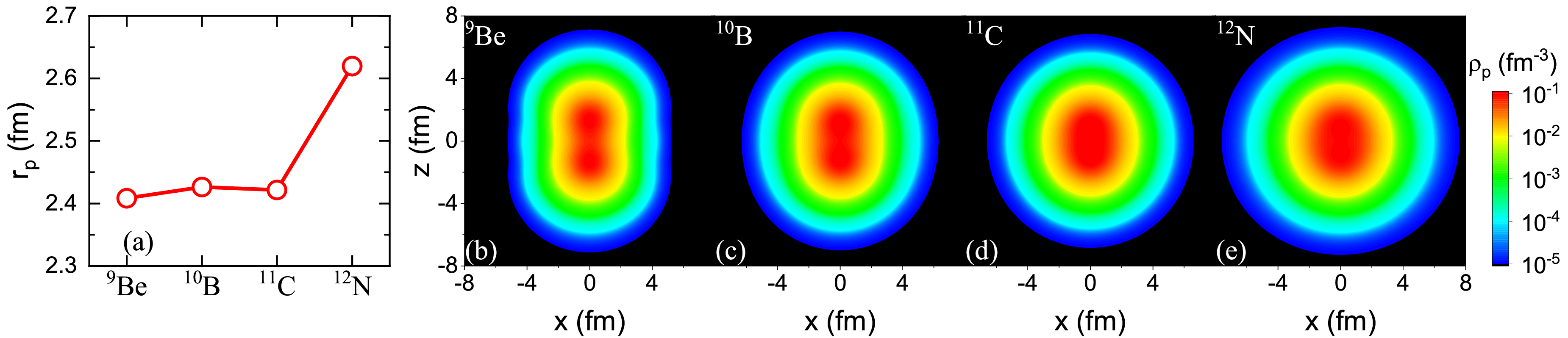}
  \caption{(a) Proton rms radii for the $N=5$ isotones from $Z=4$ ($^9$Be) to $Z=7$ ($^{12}$N), and (b--e) their proton density distributions in the $xz$ plane with $z$ being the symmetry axis.}
\label{fig2}
\end{figure*}

As evident from Fig.~\ref{fig1}, the DRHBc results are robust and consistent across the adopted density functionals and pairing strengths. Therefore, unless otherwise specified, the following discussion will focus on the results obtained with the NL3$^*$ functional and a pairing strength of $-374$~MeV~fm$^3$ \cite{Li2012PRC}.
To visualize the proton halo structure, Fig. \ref{fig2} presents the proton rms radii $r_p$ and two-dimensional proton density distributions for the $N=5$ isotones.
As shown in Fig. \ref{fig2}(a), the proton radii remain nearly constant from $^9$Be to $^{11}$C, suggesting that the increase in proton number is almost offset by the decreasing deformation.
In contrast, the $r_p$ of $^{12}$N shows a sudden increase, deviating dramatically from the trend.
In Figs. \ref{fig2}(b)--\ref{fig2}(d), the proton density distributions exhibit progressively reduced elongation along the symmetry axis $z$, consistent with the decreasing prolate deformation. Notably, a two-center cluster structure is visible in $^9$Be, corresponding to its large deformation shown in Fig. \ref{fig1}(c).
This is consistent with interpretations based on the reaction dynamics of $^9$Be+$^{110}$Pd collisions~\cite{Kundu2025PLB} and with the deformation $\beta_2 = 0.755^{+0.11}_{-0.11}$ extracted from $\alpha$ scattering experiments~\cite{Roy1995PRC}.
In Fig. \ref{fig2}(e), the proton density distribution of $^{12}$N reveals a pronounced spatial extension, particularly in the direction perpendicular to the symmetry axis.
This explains the sudden increase in $r_p$ and reveals a shape decoupling between the core and the halo in $^{12}$N, where a spatially extended oblate proton halo surrounds a prolate core--a feature that will be discussed in more detail below.

\begin{figure}[H]
  \centering
  \includegraphics[width=0.45\textwidth]{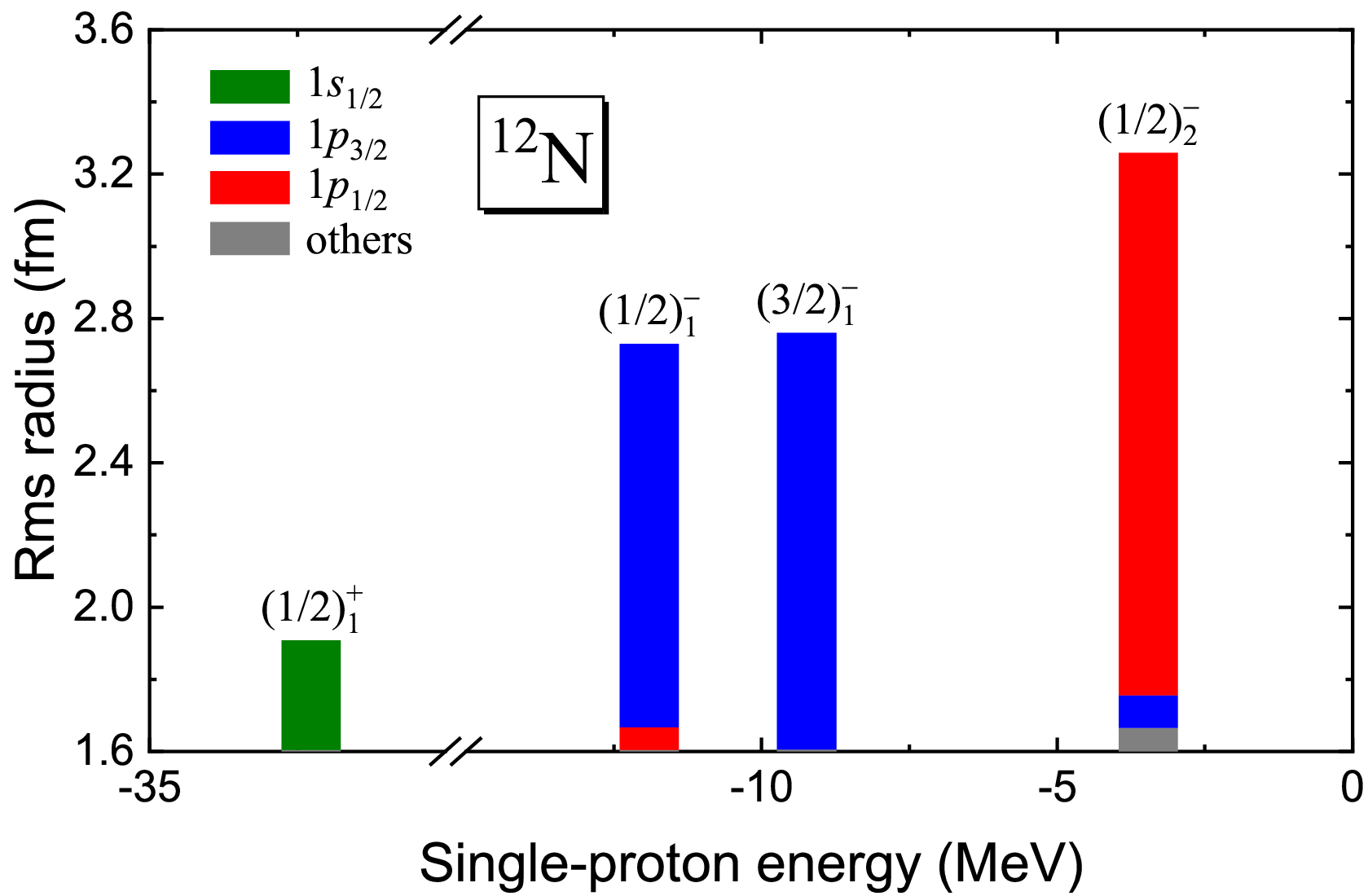}
  \caption{Rms radii versus single-particle energies for the proton orbitals in the canonical basis for $^{12}$N. Each orbital, labeled by $\Omega^\pi_i$, is represented by a vertical bar, with different colors indicating the contributions from its main components.}
\label{fig3}
\end{figure}

To understand the microscopic origin of the halo in $^{12}$N, we examine the single-proton levels in the canonical basis, along with their rms radii and component decompositions, as shown in Fig. \ref{fig3}.
Each level is labeled by $\Omega^\pi_i$, where $\Omega$ denotes the projection of the total angular momentum onto the symmetry axis, $\pi$ the parity, and $i$ the ordinal index distinguishing levels with the same $\Omega^\pi$.
Among all orbitals, the weakly bound $(1/2)_2^-$ orbital occupied by the valence proton stands out with an exceptionally large rms radius exceeding 3.2 fm, significantly larger than those of the other proton orbitals and beyond the matter rms radius of the nucleus.
This dramatic spatial extension arises from the combined effects of weak binding and dominant $p$-wave components, which together facilitate substantial wavefunction tunneling into the classically forbidden region \cite{Zhang2019PRC}.
In contrast, although other orbitals also feature low-$\ell$ components, their spatial extension is suppressed by deeper binding.
Interestingly, the single-proton spectrum of $^{12}$N ($Z=7$) bears a resemblance to the single-neutron spectrum of the classic halo nucleus $^{11}$Li ($N=8$) in the RCHB description \cite{Meng1996PRL}.
However, $^{11}$Li features a larger isospin asymmetry, and its valence neutrons are not subject to the additional binding introduced by the Coulomb barrier.
In addition, the unpaired proton in $^{12}$N induces blocking effects that suppress pairing correlations.
In contrast, the stronger pairing in $^{11}$Li can scatter its more weakly bound valence neutrons into the continuum $2s_{1/2}$ orbital \cite{Meng1996PRL}, leading to a more prominent halo than in $^{12}$N.
It is also noteworthy that in an extended three-body model, pairing correlations drive the halo formation in $^{11}\mathrm{Li}$ by mixing the $(2s)^2$ and $(1p)^2$ configurations with nearly equal weight \cite{Myo2007PRC}.

\begin{figure}[H]
  \centering
  \includegraphics[width=0.4\textwidth]{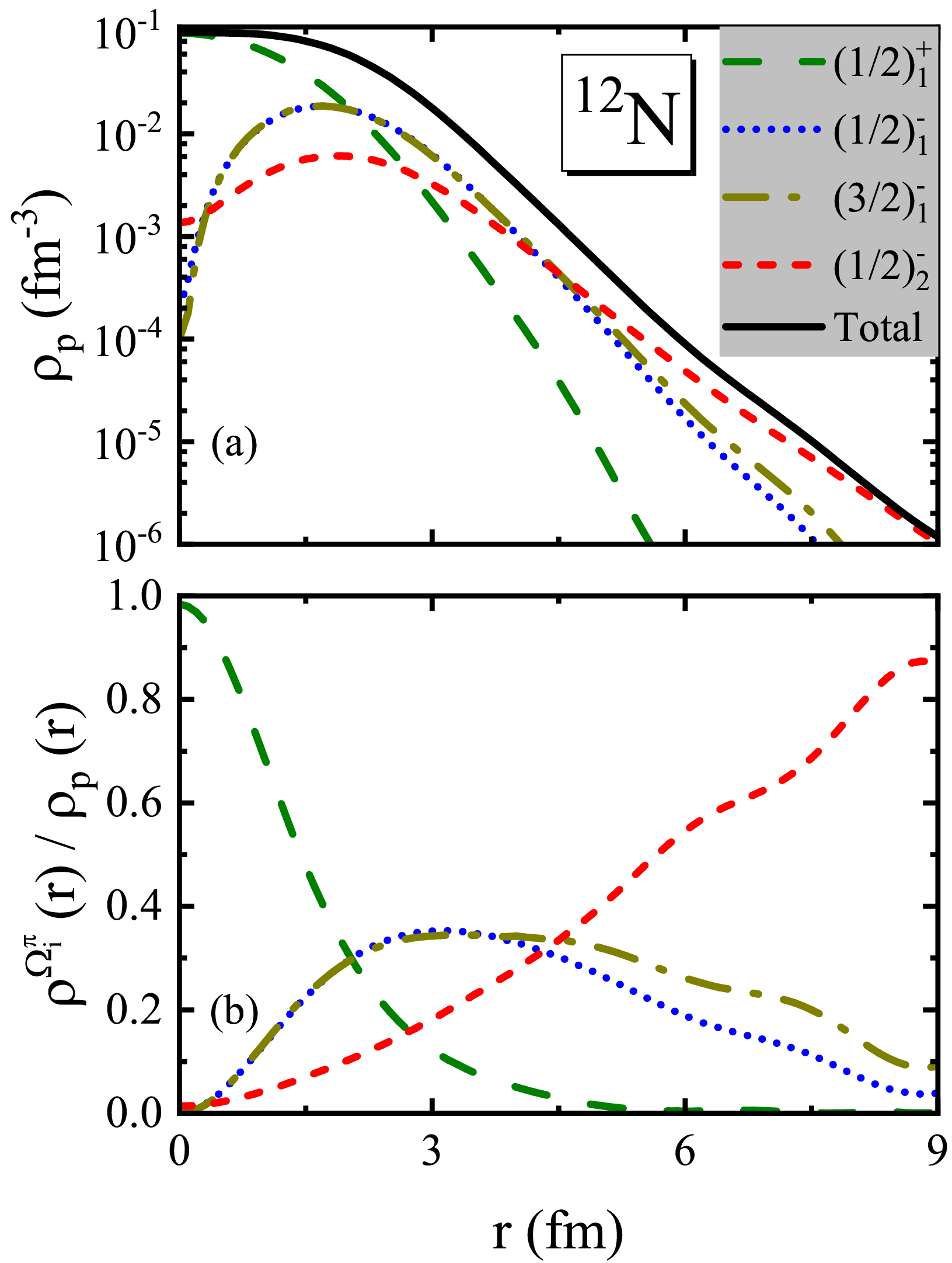}
  \caption{(a) Angle-averaged densities of individual single-proton orbitals in $^{12}$N, along with the total proton density, plotted as functions of the radial coordinate $r$; (b) Corresponding contributions of each orbital to the total proton density as functions of $r$.}
\label{fig4}
\end{figure}

To further elucidate the role of the valence proton in halo formation, the angle-averaged proton density and the contributions from individual orbitals are displayed in Fig.~\ref{fig4}.
In the central region ($r \lesssim 2$ fm), the density is dominated by the most deeply bound $s$-wave orbital.
In the intermediate region ($2 \lesssim r \lesssim 4.5$ fm), all four orbitals make appreciable contributions to the total proton density.
Beyond this range, the valence proton contribution becomes increasingly dominant.
In particular, its contribution rises steadily with $r$, reaching about $90\%$ at large distances.
This clearly highlights the decisive role of the valence proton in driving the spatial extension of the proton density in $^{12}$N, consistent with the classical picture of nuclear halos.
Therefore, a fully microscopic, self-consistent description of the proton halo phenomenon in $^{12}$N is achieved by the DRHBc theory.

\begin{figure}[htbp]
  \centering
  \includegraphics[width=0.45\textwidth]{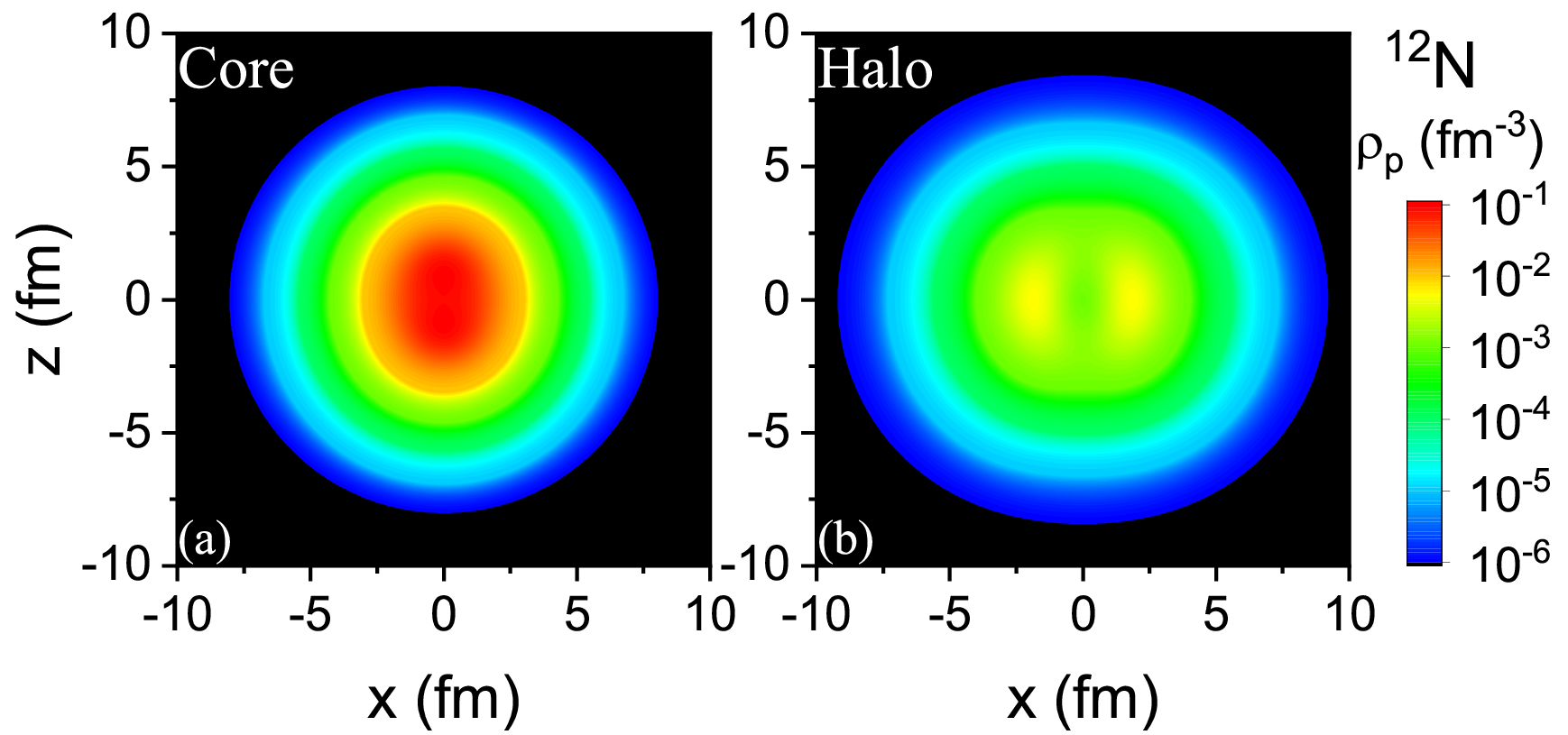}
  \caption{Same as Fig. \ref{fig2}(e), but shown separately for (a) the core and (b) the halo of $^{12}$N.}
\label{fig5}
\end{figure}

As shown in Figs. \ref{fig3} and \ref{fig4}, the valence proton orbital is clearly decoupled from the others in both energy and spatial distribution.
This enables the total proton density of $^{12}$N to be naturally decomposed into core and halo contributions, as illustrated in Figs. \ref{fig5}(a) and \ref{fig5}(b), respectively.
The prolate core exhibits a density pattern similar to the proton density of $^{11}$C shown in Fig. \ref{fig2}(d).
Quantitatively, it has an rms radius of $r_p = 2.497$ fm and a quadrupole deformation of $\beta_2 = 0.198$, both close to the corresponding values in $^{11}$C.
In contrast, the halo density extends much farther in the $xy$ plane perpendicular to the symmetry axis, and features a markedly larger rms radius of $r_p = 3.259$ fm and an oblate deformation characterized by $\beta_2 = -0.325$.
The shape decoupling between the core and the halo, such as spherical-prolate \cite{Pan2024PLB}, prolate-oblate \cite{Li2012PRC}, higher-order \cite{Zhang2023PLB}, and triaxial \cite{Zhang2023PRC(L2)} deformation decouplings, has become a well-established concept in theoretical studies of deformed halo nuclei, where the halo deformation is primarily governed by the wave function of valence nucleon(s) \cite{Zhou2010PRC(R),Misu1997NPA}.
In $^{12}$N, the halo proton occupies the $(1/2)^-_{2}$ orbital, composed of over $90\%$ $1p_{1/2}$ component, as shown in Fig.~\ref{fig3}.
The angular momentum coupling in this case is analogous to that in the predicted deformed halo nuclei $^{39}$Na \cite{Zhang2023PRC(L1)} and $^{42,44}$Mg \cite{Zhou2010PRC(R),Li2012PRC}, where valence neutrons partially occupy $(1/2)^-$ orbitals with $2p$ components.
This results in an oblate halo, whose angular distribution is predominantly dictated by the spherical harmonics $|Y_{1\pm1}|^2 \propto \sin^2\theta$.

\section{Summary}\label{summary}

In conclusion, the deformed relativistic Hartree-Bogoliubov theory in continuum has been applied for the first time to investigate the proton halo phenomenon.
A fully microscopic, self-consistent description of the proton halo in $^{12}$N is achieved.
Ground-state properties of the $N=5$ isotones, from the stable nucleus $^9$Be to the proton drip-line nucleus $^{12}$N, have been studied.
The DRHBc calculations reproduce the available experimental proton separation energies and empirical matter rms radii reasonably well, without introducing any adjustable parameters.
From $^9$Be to $^{11}$C, as the proton number increases, the proton rms radius remains nearly unchanged while the matter rms radius gradually decreases, primarily due to the reduction in nuclear deformation.
However, from $^{11}$C to $^{12}$N, despite a further decrease in deformation, both proton and matter rms radii exhibit a striking enhancement, revealing the emergence of a proton halo in $^{12}$N, as evidenced by its extended proton density distribution.
This halo originates from the occupation of a weakly bound valence orbital dominated by $p$-wave components, which enables significant wavefunction tunneling into the classically forbidden region and results in a substantially larger rms radius than those of the core orbitals.
The contribution of the halo proton to the total proton density increases steadily with radial distance, reaching approximately $90\%$ at large $r$.
While the core exhibits a prolate shape elongated along the symmetry axis, the halo density is mainly distributed in the perpendicular direction, indicating a well-defined shape decoupling between the prolate core and the oblate halo in $^{12}$N.

\section*{Acknowledgments}

We extend our gratitude to the anonymous referee for their meticulous and constructive feedback, which has significantly improved this work.
We also thank members of the DRHBc Mass Table Collaboration for helpful discussions, especially C. Pan and S. Q. Zhang for their valuable insights.
This work was partly supported by the National Natural Science Foundation of China (Grant No. 12305125), the National Key Laboratory of Neutron Science and Technology (Grant No. NST202401016), and Sichuan Science and Technology Program (Grant No. 2024NSFSC1356).


\end{document}